\documentclass[aps,prl,twocolumn,superscriptaddress]{revtex4-1}
\usepackage{graphicx,bm,amssymb,amsmath,xcolor,pifont,soul}
\usepackage[linktocpage=true,colorlinks=true,pdfborder={0 0 0},linkcolor=blue,citecolor=red,filecolor=yellow,urlcolor=blue,bookmarks,pdfauthor={},]{hyperref}
\usepackage{ulem}
\usepackage{multirow}
\usepackage[T1]{fontenc}
\newcommand{\HP}[1]{\textcolor{black}{#1}}

\begin{document}

\title{Implications of the electron-phonon coupling in CuPb$_9$(PO$_4$)$_6$O for superconductivity: an \textit{ab initio} study}
\author{Hari Paudyal}
\email{hari-paudyal@uiowa.edu}
\affiliation{Department of Physics and Astronomy, University of Iowa, Iowa City, Iowa 52242, USA}
\affiliation{Ames National Laboratory of the US DOE, Iowa State University, Ames, IA 50011, USA}
\author{Michael E. Flatt\'e}
\email{michaelflatte@quantumsci.net}
\affiliation{Department of Physics and Astronomy, University of Iowa, Iowa City, Iowa 52242, USA}
\affiliation{Department of Applied Physics, Eindhoven University of Technology, Eindhoven, The Netherlands}
\author{Durga Paudyal}
\email{durga@ameslab.gov}
\affiliation{Ames National Laboratory of the US DOE, Iowa State University, Ames, IA 50011, USA}
\affiliation{Department of Electrical and Computer Engineering, Iowa State University, Ames, IA 50011, USA}
\date{\today}

\begin{abstract}
We report $ab~initio$ calculations of the electronic and vibrational properties in CuPb$_9$(PO$_4$)$_6$O, including the electron-phonon coupling strength via strong-coupling Migdal-Eliashberg theory.  We verify the presence of appealing flat electronic bands near the Fermi level, a strong hybridization between the Cu $3d$ and O $2p$ states, and soft low-energy phonons, which can suggest high-temperature superconducting behavior. However, the electron-phonon coupling strength appears insufficient to overcome the Coulomb repulsion between an electron pair and thus does not support high-temperature superconductivity in CuPb$_9$(PO$_4$)$_6$O via the conventional electron-phonon Migdal-Eliashberg mechanism. Even neglecting Coulomb repulsion of the electron pair we find this electron-phonon coupling suggests a superconducting transition temperature less than 2~K.

\end{abstract}
\maketitle

Ambient pressure superconductivity at room temperature has recently been suggested in Cu$_x$Pb$_{10-x}$(PO$_4$)$_6$O~\cite{Lee_arxiv23First, Lee_arxiv23Supercond}, indicating possible progress towards this challenging and  elusive goal~\cite{Flores_PhysRep20perspective, Lilia_JPCM22}. In the last decades MgB$_2$~\cite{Nagamatsu_Nat01MgB2} has advanced superconducting technologies 
\HP{due to the ability to form and process MgB$_2$ into useful forms, despite a critical temperature~\cite{Braccini_PhysC07MgB2, Penco_IEEE07MgB2} far from the maximum achieved in a material}. Compacted superhydrides have exhibited conventional superconducting behavior at elevated temperatures under extreme pressure~\cite{Drozdov_Nat15H3S, Somayazulu_PRL19, Kong_NatCommun21}, however, employing such pressure limits these materials' potential utility. Therefore, the quest for room temperature ambient pressure superconductors represents a groundbreaking pursuit and its achievement would revolutionize  applications including energy transmission and storage, transportation, and healthcare~\cite{Yao2021superconducting_applications}. 
Recently Cu$_x$Pb$_{10-x}$(PO$_4$)$_6$O with $0.9<x<1.1$ has  exhibited  a sudden drop in resistivity, along with a negative diamagnetic susceptibility, and distinct voltage spikes below 400~K, attributed to the strain due to the Cu doping~\cite{Lee_arxiv23First, Lee_arxiv23Supercond}, leading to suggestions it might exhibit high-temperature superconductivity.  
A  solid-state reaction recipe was presented for the synthesis of this material, however, the experimental characterization only measures an average crystal structure and cannot detect possible changes in local geometries with respect to the partly filled O sites. The absence to date of independent reproduction of these results~\cite{Liu_arxiv23_expt, Kumar_arxiv23, Kumar_Super23}  requires  further experimental validation, and suggests that theoretical studies may be helpful. Some electronic structure calculations~\cite{Kumar_arxiv23, Si-Held_arxiv23, Si-Simone_arxiv23, Sun-Vlamidir_arxiv23, Lai_JMST23} provide a few encouraging but limited results. The appearance of flat CuO bands, with a relatively high density of states (DOS) at the Fermi level ($E_{\rm F}$), could support  high-temperature superconductivity. However, those calculations do not rule out  magnetic configurations which could suppress the superconducting behavior~\cite{Sun-Vlamidir_arxiv23, Zhang_arxiv23, Das2022s}. A few studies suggest substantial effective electron-phonon coupling due to this elevated DOS in the proximity to  $E_{\rm F}$ in this material~\cite{Griffin_arxiv23, Cabezas_arxiv23}, but did not directly simulate the electron-phonon coupling.



In this letter, we report a systematic study of the electronic, vibrational, and electron-phonon coupling properties of CuPb$_9$(PO$_4$)$_6$O through \textit{ab initio} calculations within strong-coupling electron-phonon superconductivity (Migdal-Eliashberg) theory. The calculated lattice parameters are found in good agreement with  previously reported results, as well as confirming that lattice compression  occurs with Cu substitution. The insulating to metallic transition is attributed to the presence of flat Cu $3d$ states hybridized with O $2p$ states at  $E_{\rm F}$. As a result, our electronic structure calculations exhibit a relatively large DOS at  $E_{\rm F}$, and multiple Van Hove singularities~\cite{Sun-Vlamidir_arxiv23}. Phonon calculations show that  CuPb$_9$(PO$_4$)$_6$O  is dynamically stable with many low frequency vibration modes below 5~meV. Despite having a high DOS at  $E_{\rm F}$ and low energy phonons, we found a weak electron-phonon coupling strength that would only provide pairing below 2K for conventional superconductivity in this material. 

Our \textit{ab initio} calculations have been performed using \texttt{Quantum ESPRESSO}~\cite{giannozzi2017o} with  relativistic norm-conserving pseudopotentials~\cite{PseudoDojo2018} in conjunction with the
Perdew-Burke-Ernzerhof~\cite{perdew1996o} exchange-correlation functional in the generalized gradient approximation. A plane wave kinetic-energy (charge density) cutoff value of 80~Ry (320~Ry), a $\Gamma$-centered $8\times 8 \times 8$ Monkhorst-Pack \textbf{k}-mesh~\cite{Monkhors1976s}, and a Methfessel and Paxton smearing~\cite{Methfessel1989s} width of 0.01~Ry have been used for the Brillouin-zone sampling. The atomic positions and lattice parameters are optimized until the self-consistent energy is converged within $2.7\times10^{-5}$ eV and the maximum force on each atom is less than 0.005~eV/\AA. For the DOS and Fermi surface calculations, denser  \textbf{k}-meshes of $16 \times 16 \times 16$ and $32 \times 32 \times 32$ are used. The dynamical matrices and the linear variation of the self-consistent potential are calculated within density-functional perturbation theory~\cite{Baroni2001s} on an irreducible $4 \times 4 \times 4$ \textbf{q} mesh. 

\begin{figure}[t]
\centering	
\includegraphics[width=0.45\textwidth]{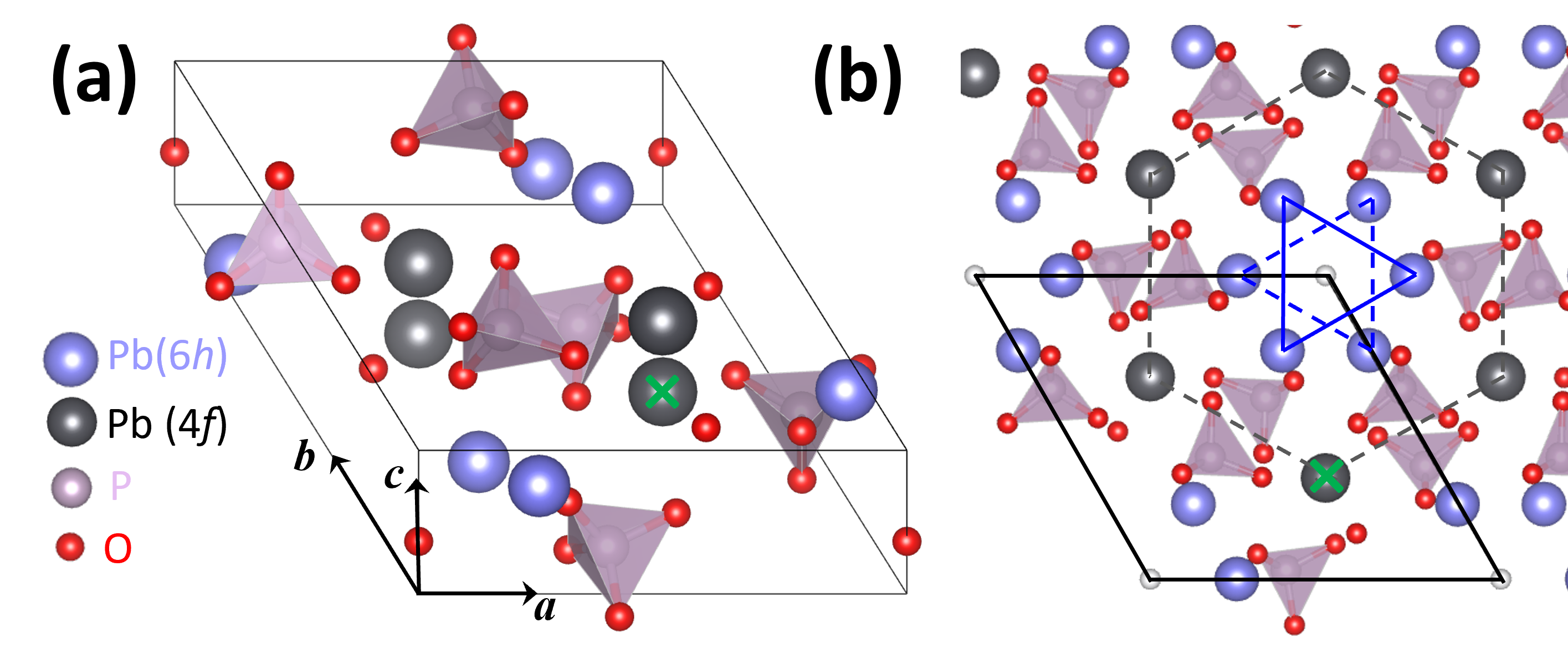}\hfill	
\caption{Crystal structure of Pb$_{10}$(PO$_4$)$_6$O in a (a) single unit cell and (b) supercell. The O with 25\% occupancy lies outside of the PO$_4$ tetrahedra. There are two distinct non-equivalent Pb($6h$) and Pb($4f$) sites. The Pb($6h$) atoms form two triangular layers (with 3.71 \AA~vertical separation and 60$^{\rm o}$ rotation), whereas the Pb($4f$) atoms adopt a hexagonal arrangement. For Cu substitution we considered the Pb($4f$) site marked with a green cross.}
\label{cryst-str}
\end{figure}

 Migdal-Eliashberg theory~\cite{Allen1975s,Margine2013s}, as implemented in the Electron-Phonon Wannier (\texttt{EPW}) code~\cite{ponce2016o, EPW2023}, is used to investigate the electron-phonon properties. 
 Wannier interpolation~\cite{Pizzi2019s} on a uniform $8 \times 8 \times 8$ $\Gamma$-centered \textbf{k}-grid is used to calculate the Cu $3d$ and O $2p$ orbital Wannier functions used in the electron-phonon calculations. Denser (uniform) $20\times 20 \times 20$ \textbf{k}- and \textbf{q}-point grids are used for electron-phonon matrix element calculations. Effective Coulomb repulsion values of $\mu^* = 0, 0.1$, $0.15$, and $0.2$ are used to estimate the superconducting critical temperature by self-consistently solving the isotropic Migdal-Eliashberg equations. 
 The Matsubara frequency cutoff is set to 1.5~eV and the Dirac deltas are replaced by Lorentzians of width 50~meV for electrons and 0.1~meV for phonons.

The apatite crystal structure of Cu$_x$Pb$_{10-x}$(PO$_4$)$_6$O with 0.9 $<$ x $<$ 1.1 is obtained by replacing a specific Pb sublattice with a Cu atom from the parent Pb$_{10}$(PO$_4$)$_6$O compound~\cite{krivovichev2003crystal}. The parent structure [Fig.~\ref{cryst-str}(a)], which exhibits semiconducting behavior~\cite{Cabezas_arxiv23, Si-Held_arxiv23, Liu_arxiv23phonon}, contains six Pb($6h$) and four Pb($4f$) sites as well as six PO$_4$ tetrahedra each containing P, O(1), O(2), and O(3) with $6h$, $6h$, $12i$, and $6h$ sites, respectively, forming a unit cell with hexagonal symmetry. \HP{We notice that Pb(6$h$) and Pb(4$f$) are often inconsistently labeled as Pb(1) and Pb(2) in the literature~\cite{krivovichev2003crystal, kurleto2023pb, Si-Held_arxiv23, Liu_arxiv23phonon}.} The Pb($6h$) atoms form two triangular layers connected by a hexagonal symmetry plane perpendicular to the $c$-axis and separated by 3.71~\AA~with a 60$^{\rm o}$ rotation, whereas the Pb($4f$) atoms adopt a hexagonal arrangement [Fig.~\ref{cryst-str}(b)]. In addition, four O(4) atoms occupy $4e$ sites constituting a quarter of the occupancy [Fig.~\ref{cryst-str}(a)]. \HP{We note that in Ref.~\onlinecite{Lee_arxiv23First} (Fig. 3e) oxygens outside the unit cell are shown, which can suggest that there are two O(4) atoms; only one O(4) atom is in each cell~\cite{krivovichev2003crystal, kurleto2023pb, Si-Held_arxiv23, Liu_arxiv23phonon}.} To model this partial occupancy we considered the structure with the lowest energy, simultaneously maintaining  full occupation of  one out of the four O(4) sites. Here, the energy difference between different oxygen occupancies is less than 0.2~meV/atom. The structural optimization slightly shifts the position of the O(4) atom along the $c$-axis. This O atom lies outside the PO$_4$ tetrahedra, inducing a potential local distortion.

\begin{figure}[!t]
\centering	
\includegraphics[width=0.45\textwidth]{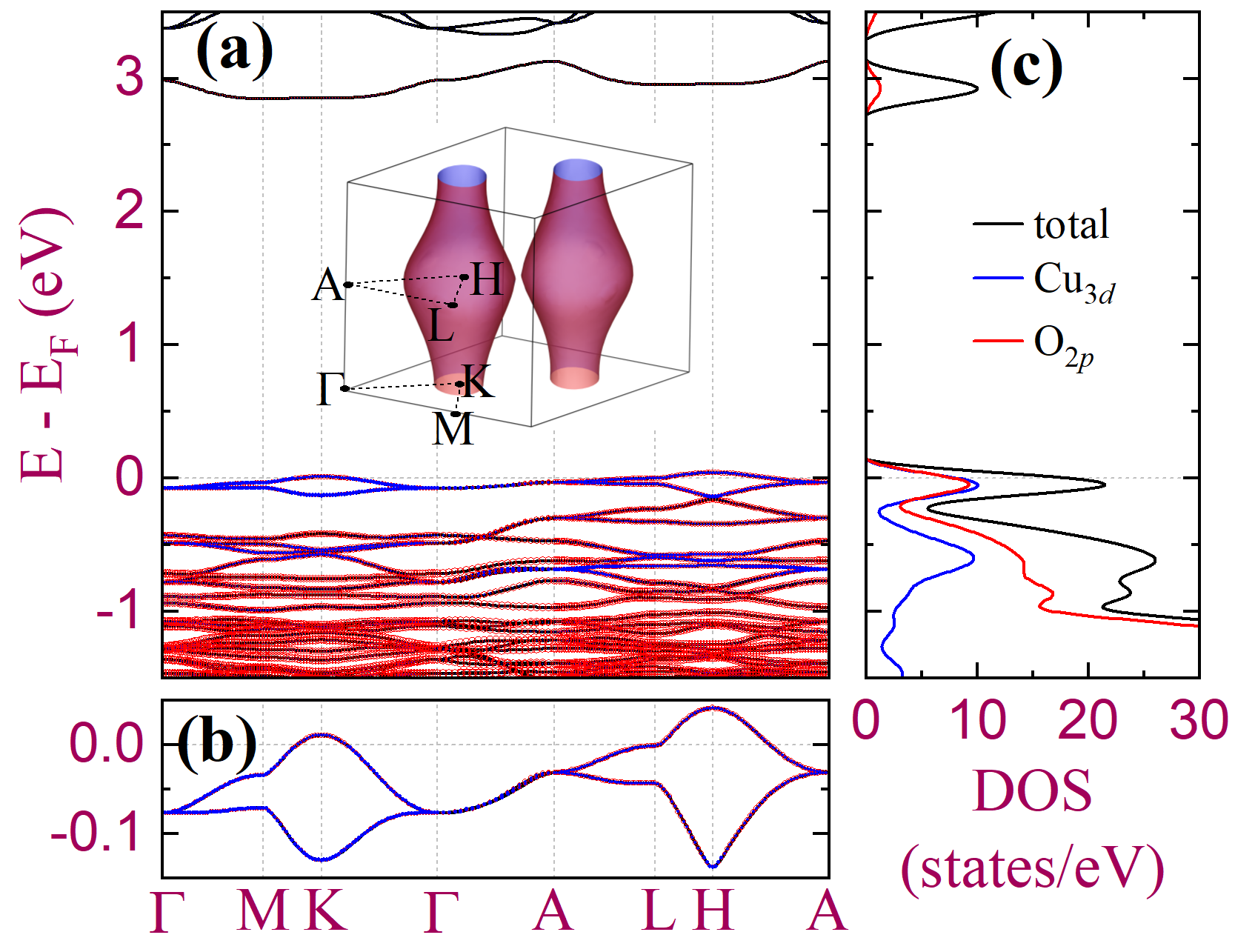}\hfill	
\caption{Electronic properties of CuPb$_{9}$(PO$_4$)$_6$O: band structure in wide (a) and narrow (b) energy windows, density of states (c), and Fermi surface (inset)~\cite{Kawamura2019s}.}
\label{band-dos}
\end{figure}

To model the CuPb$_9$(PO$_4$)$_6$O structure we replaced one of the Pb atoms by Cu. However, the exact site within the Pb sublattice where the Cu atom is substituted remains uncertain. Therefore we examined Cu substitutions in one of the (identical) Pb($6h$) and one of the (identical) Pb($4f$) sites individually to identify the position of lowest energy. These calculations confirm that a Cu at the Pb($6h$) site is energetically favorable forming a bond of length 1.93~\AA~with Cu-O(4). Instead of transforming the system to metallic, the structure remains semiconducting with a band gap of $\sim$1~eV, consistent with previous results~\cite{Zhang_arxiv23, Liu_arxiv23phonon}. The structure with Cu \HP{substituted} at the Pb($4f$) site is  $\sim$0.3~eV/f.u (per formula unit) higher in energy with respect to the one substituted at the Pb($6h$) site, in agreement with previous work~\cite{Sun-Vlamidir_arxiv23}, and shows a metallic electronic band structure (discussed later). As Migdal-Eliashberg theory begins from a metallic state, 
we assumed the Pb($4f$) location was preferred over the Pb($6h$) site, as shown by the green cross in Fig.~\ref{cryst-str}(a). The optimized lattice parameters of the Cu-substituted system ($a$ = 9.957 \AA, $c$ = 7.430 \AA, and $V$ = 637.94 \AA$^3$) decrease relative to the parent structure ($a$ = 10.032 \AA, $c$ = 7.492 \AA, and $V$ = 652.99 \AA$^3$). Here, the position of O(4) is found displaced by $\sim$0.3 \AA~along the $c$-axis without altering the hexagonal symmetry. The shrinkage of lattice parameters and reduction in volume agrees with  experimental and theoretical reports~\cite{Lee_arxiv23First, Si-Held_arxiv23}.

\begin{figure*}[!ht]
\centering	
\includegraphics[width=0.99\textwidth]{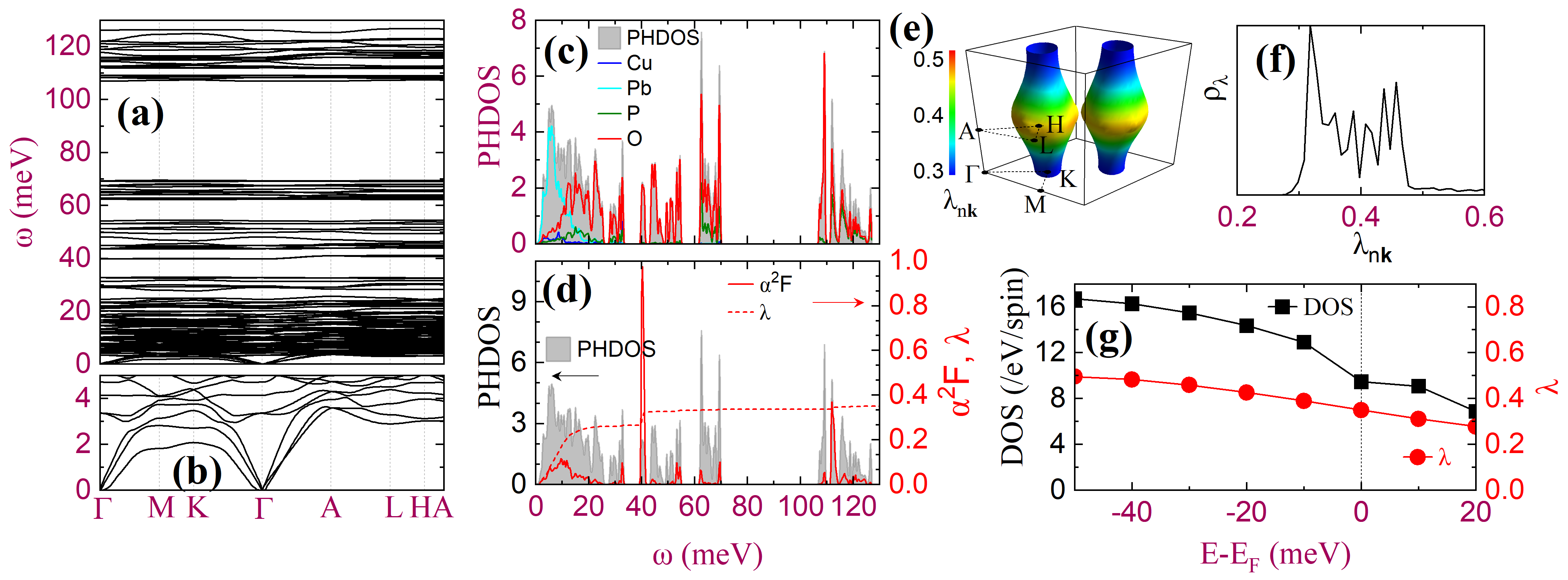}\hfill	
\caption{Vibrational and electron-phonon properties of CuPb$_{9}$(PO$_4$)$_6$O. Phonon dispersions are plotted in wide (a) and narrow (b) energy windows.  (c) Phonon density of states (PHDOS) along with the atomic projection of the PHDOS.  (d) PHDOS, Eliashberg spectral function $\alpha^2F(\omega)$, and total electron-phonon coupling strength $\lambda$, (where $\lambda(\omega)=\int\left[\alpha^2F(\omega)/(2\omega)\right]d\omega)$ and the units of $\alpha^2F(\omega)$ are normalized to its peak value). (e) Momentum-resolved anisotropic electron-phonon coupling $\lambda_{n\textbf{k}}$ on the Fermi surface plotted in a primitive Brillouin zone~\cite{Kawamura2019s}.  (f) Distribution of the electron-phonon coupling strength $\lambda_\textbf{k}$.  (g) Variation of the DOS and $\lambda$ with respect to energy relative to $E_{\rm F}$.}
\label{phonon}
\end{figure*}

Figure~\ref{band-dos}(a) and \ref{band-dos}(b) show the electronic band structure of CuPb$_9$(PO$_4$)$_6$O along the high symmetry points of the full Brillouin zone. The Brillouin zone is hexagonal and the two identical sheets of the Fermi surface lie on the identical K and K' points, as shown in the inset of Fig.~\ref{band-dos}(a). The Cu $3d$ and O $2p$ orbital contributions are indicated by the blue and red colors. Here, the 2$p_x$ and 2$p_y$ orbitals of O strongly hybridize with  all the 3$d$ orbitals of Cu, except 3$d_{z^2}$. Fig.~\ref{band-dos}(b) shows the metallic band structure  within a small energy window close to the $E_{\rm F}$, in which a flat band, predominantly of Cu $3d$ and O $2p$ character, crosses  $E_{\rm F}$. This band has a camel shape dispersion with the two humps at the K and H points of the Brillouin zone along the M-K-$\Gamma$ and L-H-A directions. These crossings produce a waisted cylindrical shaped Fermi surface with a K (or K$^\prime$)-centered hole pocket [Fig.~\ref{band-dos}(a) inset]. The DOS and orbital projections are  shown in Fig.~\ref{band-dos}(c). The Cu $3d$ and O $2p$ orbitals contribute equally at  $E_{\rm F}$. As the electronic band structure and DOS remain relatively unchanged across a broad energy range, spin-orbit coupling is not considered. We note that the electronic structure of CuPb$_9$(PO$_4$)$_6$O is not consistent among the recent theoretical studies~\cite{Si-Held_arxiv23, Si-Simone_arxiv23, Kumar_arxiv23}, especially regarding the relative position of  $E_{\rm F}$. The primary reason for this inconsistency appears to be sensitivity to the convergence parameters (such as \textbf{k}-mesh, energy cut-off, and the value of smearing) and the pseudopotential types used. We have used  norm-conserving  pseudopotentials and found that the relative position of  $E_{\rm F}$ lies $\sim$30~meV above  the one calculated with projector augmented wave pseudopotentials, especially in the $\Gamma$-A Brillouin zone direction. Therefore, we must examine the sensitivity of our results (discussed in the electron-phonon coupling) to a small shift on either side of $E_{\rm F}$.

Figure~\ref{phonon}(a), \ref{phonon}(b), and \ref{phonon}(c) show the calculated phonon dispersions of CuPb$_9$(PO$_4$)$_6$O along with the atom-projected phonon DOS (PHDOS). The optical phonons exhibit a distinct flat dispersion relation across the entire vibrational spectrum, whereas the acoustic branches display conventional linear behavior. As a result of the flat phonon dispersion, sharp peaks in the PHDOS are observed. The complete spectrum is divided into four parts: low frequency vibrations below 35~meV, intermediate frequency vibrations 40-55~meV and 60-70~meV, and high frequency vibrations above 105~meV. The lower frequency modes predominantly correspond to  vibrations of the Pb atoms, whereas the modes within the intermediate range (60-70~meV) and the higher frequency modes are associated with  vibrations of the P atoms. The vibrations of the O atoms extend across the entire spectrum; they are an integral part of the PO$_4$ tetrahedra, giving rise to  collective oscillations involving mixed vibrational modes with other atoms. This indicates a complex interplay of chemical bonding and structural flexibility of the PO$_4$ tetrahedra in CuPb$_9$(PO$_4$)$_6$O. Even within the harmonic approximation the Cu substitution primarily impacts the low-frequency modes below 15~meV [Fig.~\ref{phonon}(c)], as Cu is considerably heavier than the O and P atoms. Cu vibrations, mixing with the lighter atom modes, may induce substantial anharmonicity and mode coupling, highlighting the pivotal role of lattice distortions on the electronic and vibrational properties.

To further assess whether CuPb$_9$(PO$_4$)$_6$O would superconduct via the strong-coupling electron-phonon mechanism, we examine the isotropic Eliashberg spectral function $\alpha^2F(\omega)$ and the cumulative electron-phonon coupling strength $\lambda$ [Fig.~\ref{phonon}(d)]. A sharp increase in the $\alpha^2F(\omega)$ distribution is found up to 20~meV, indicating strong electron-phonon coupling in the low frequency region. This is unsurprising  because of the large value of the DOS at  $E_{\rm F}$. 
Further, near the intermediate frequency 40~meV a sharp peak is noticed in  $\alpha^2F(\omega)$. A continuous isotropic electron-phonon coupling strength $\lambda_{n\textbf{k}}$, ranging from 0.3 to 0.5, develops on the Fermi surface [Fig.~\ref{phonon}(e) and Fig.~\ref{phonon}(f)] with a peak at the K-point of the Brillouin zone. Consequently, a weak ($\sim$0.4) average electron-phonon coupling strength is found, insufficient to support conventional electron-phonon strong-coupling  high-temperature superconductivity. It is concerning that the obtained value of $\lambda$ lies within the upper range of 
the Coulomb interaction parameters that have been reported for some conventional superconductors~\cite{an2021superconductivity, Heil2017s, Kostrzewa_SciRep18_Coulomb}; thus the electron-phonon pairing is unlikely to be robust in the presence of repulsion, implying the superconducting transition temperature will not exceed a few K. 

Even neglecting the Coulomb repulsion of the electron pair we find that the value we obtain for the electron-phonon coupling suggests a superconducting transition temperature less than 2~K. We estimate the superconducting transition temperature by solving temperature dependent coupled nonlinear Eliashberg equations to find the superconducting gap, incorporating the effective Coulomb repulsion $\mu^*$, in the isotropic approximation~\cite{Allen-Mitro1983, Margine2013s}. The temperature at which the gap vanishes is the superconducting transition temperature. Calculated superconducting transition temperatures of 250~mK, 26~mK, and $\sim$0~mK correspond to more realistic  effective Coulomb repulsion values of $\mu^* = 0.1$, $0.15$, and $0.2$. Hence we do not see any signature of  high-temperature superconductivity in CuPb$_9$(PO$_4$)$_6$O  that can be described by  conventional strong-coupling Migdal-Eliashberg theory.

To assess the accuracy of these  electron-phonon calculations for CuPb$_9$(PO$_4$)$_6$O, we examined their sensitivity to the Fermi energy. The variation of the DOS and the cumultative electron-phonon coupling strength are shown in Fig.~\ref{phonon}(g) for a rigid shift in the Fermi energy on either side of  $E_{\rm F}$ between -50~meV and 20~meV in steps of 10~meV.  As expected the value of electron-phonon coupling strength increases as the DOS increases (and conversely), and reaches its largest value, 0.5, for a -50~meV shift in  $E_{\rm F}$. The overall shape of  $\alpha^2F(\omega)$ for a shifted Fermi energy  is similar to the one calculated at the actual $E_{\rm F}$, but with a stronger coupling observed in the low-frequency modes. However, the value of the electron-phonon coupling strength remains insufficient to overcome  electron-electron repulsion and produce high-temperature superconductivity.

To understand the correlated electron magnetism in CuPb$_9$(PO$_4$)$_6$O, we have performed spin polarized electronic structure calculations incorporating electron-electron correlation (Hubbard parameter $U_{eff}=U-J$ = 3.0~eV~\cite{Si-Held_arxiv23} where $U$ and $J$ are the \HP{on-site} Coulomb repulsion and on-site exchange interaction). Both with and without $U_{eff}$ we found a total magnetic moment of $\sim$1~$\mu_B$/f.u. (with $\sim$0.6~$\mu_B$/Cu) in the ferromagnetic spin configuration. These results indicate that the on-site electron correlation has no significance on the potential for superconductivity, as also suggested by recent reports~\cite{Si-Held_arxiv23, Sun-Vlamidir_arxiv23}.


We employed  \textit{ab initio} Migdal-Eliashberg theory to identify whether or not the electron-phonon coupling strength found in CuPb$_9$(PO$_4$)$_6$O is sufficient to describe its high-temperature superconducting behavior via a strong-coupling electron-phonon mechanism. Electronic structure calculations indicated flat electronic bands, described by strongly hybridized Cu $3d$ and O $2p$ states, close to the Fermi level along with relatively low non-dispersive phonon frequencies, predominantly associated with vibrations of Pb. These features could suggest strong-coupling high-temperature superconductivity via the electron-phonon mechanism. Unfortunately, we calculate a low cumulative electron-phonon coupling strength. Our  obtained value may not be  enough even to overcome the Coulomb repulsion between two electrons forming a Cooper pair and produce superconductivity at any temperature. However, even if the system can become superconducting, our calculated value of the electron-phonon coupling strength does not support  high-temperature superconductivity in CuPb$_9$(PO$_4$)$_6$O via a strong-coupling electron-phonon mechanism within Migdal-Eliashberg theory.   

This work is supported by the U.S. Department of Energy, Office of Science, Office of Basic Energy Sciences under Award Number DE-SC0016379. The Ames National Laboratory is operated for the U.S. Department of Energy by Iowa State University of Science and Technology under Contract No. DE-AC02-07CH11358. We acknowledge use of the computational facilities at the University of Iowa (Argon cluster).

\bibliography{references}
\end{document}